\documentclass{JHEP3}
\usepackage{graphicx,amsmath,amssymb}
\usepackage{epsfig,multicol}
\usepackage{epsf}
\usepackage{epstopdf}
\input{epsf.sty}

\usepackage{graphicx}
\usepackage{amssymb}
\usepackage{amsmath}

\def\half{{1\over 2}}
\def\({\left(}
\def\){\right)}
\def\[{\left[}
\def\]{\right]}

\def\e{\begin{equation}}
\def\q{\end{equation}}
\def\m{\begin{eqnarray}}
\def\n{\end{eqnarray}}

\title{$g_{\rm NL}$ in the curvaton model constrained by PLANCK}
\author{Qing-Guo Huang \footnote{huangqg@itp.ac.cn}
\\\small{\em
State Key Laboratory of Frontiers in Theoretical Physics, Institute of
Theoretical Physics, Chinese Academy of Sciences, Beijing 100190,
China} }

\abstract{
As a simplest extension to the mass-term curvaton model, the curvaton model with a polynomial potential can relax the restricted constraint from PLANCK due to the non-linear dynamics of curvaton field before it decays. We find that there is still a big room for producing a large negative $g_{\rm NL}$, but not positive $g_{\rm NL}$. For example, we only need around $10\%$ ``tuning" for $-g_{\rm NL}>10^4$.
 }


\begin{document}

\section{Introduction}

Even though inflation \cite{Guth:1980zm} is an elegant model to explain the well-known puzzles in the hot big bang model, the mechanism for generating the observable anisotropies in cosmic microwave background radiation (CMBR) and forming the large-scale structure has not been well-established. In general, one may expect that there should be many light scalar fields (compared to the Hubble scale) during inflation and the quantum fluctuations of some of them finally seed the structure formation in our universe. Since the amplitude of quantum fluctuations of the light scalar field is proportional to the Hubble parameter which almost does not evolve during inflation, a nearly scale-invariant primordial density perturbation, or equivalently the spectral index of primordial curvature perturbation $n_s\simeq 1$, can be taken as a strong prediction of inflation. This prediction has been confirmed by the CMB observations. For example, the full analysis \cite{Cheng:2013iya} of pre-PLANCK data, including the 9-year Wilkinson Microwave Anisotropy Probe (WMAP) \cite{Hinshaw:2012fq}, South Pole Telescope (SPT) \cite{Story:2012wx}, Atacama Cosmology Telescope (ACT) \cite{Sievers:2013wk}, Baryon Acoustic Oscillation \cite{BAO} and $H_0$ prior from HST project \cite{HST}, implies 
\m
n_s=0.961\pm 0.007 \quad \hbox{at $68\%$ CL}, 
\label{nspre}
\n
which is almost the same as that from PLANCK \cite{Ade:2013lta,Ade:2013rta}
\m
n_s=0.9603\pm 0.0073 \quad \hbox{at $68\%$ CL}. 
\label{nsp}
\n
Inflation also predicts gravitational wave perturbation whose amplitude compared to the scalar perturbations is measured by the so-called tensor-to-scalar ratio $r$. Unfortunately the gravitational wave perturbations have not been detected and PLANCK \cite{Ade:2013lta,Ade:2013rta} sets an upper bound on it as follows 
\m
r<0.11 \quad \hbox{at $95\%$ CL}.
\label{rp}
\n

As the simplest setup of inflation, canonical single-field slow-roll inflation is governed by a canonical scalar field $\phi$ (inflaton) and the expansion rate, the Hubble parameter, is determined by the potential energy of $\phi$. During inflation inflaton slowly rolls down its flat potential in order to achieve enough e-folding number for solving the puzzles in the hot big bang model. If the cosmic structure is completely originated by the quantum fluctuations of $\phi$, it predicts that the curvature perturbation at the non-linear orders must be very small because large self-interaction of inflaton field implies a steep potential and breaks the slow-roll conditions \cite{Maldacena:2002vr}. Quantitatively a well-understood ansatz of non-Gaussianity has a local shape and the curvature perturbation $\zeta({\bf x})$ can be expanded to the non-linear orders as follows 
\m
\zeta({\bf x})=\zeta_L({\bf x})+{3\over 5}f_{\rm NL} \zeta_L^2({\bf x})+{9\over 25} g_{\rm NL} \zeta_L^3({\bf x})+\cdots \ ,
\n
where $\zeta_L({\bf x})$ denotes the linear, Gaussian part of curvature perturbation. The sizes of the non-linear order perturbations are measured by the non-Gaussianity parameters, such as $f_{\rm NL}$, $g_{\rm NL}$ and so on. Different from single-field inflation, multi-field inflation model can easily produce large local non-Gaussianity.

In this paper we focus on a well-known multi-field inflation model, namely curvaton model \cite{Enqvist:2001zp,Lyth:2001nq,Moroi:2001ct,Sasaki:2006kq,Huang:2008zj} in which the final adiabatic curvature perturbation is generated by curvaton field $\sigma$ in the radiation dominant era far after the end of inflation.  Usually a large local non-Gaussianity is expected in the curvaton model.  
Recently PLANCK data \cite{Ade:2013ydc} provides a stringent constraint on the local non-Gaussianity: 
\m
f_{\rm NL}=2.7\pm 5.8 \quad \hbox{at $68\%$ CL}. 
\label{cfnl}
\n
It implies that a large local bispectrum is unlikely. Even though the curvaton model can still fit PLANCK data well, there is no doubt that it has been tightly constrained. On the other hand, the size of trispectrum has not been constrained significantly. There are two shapes of local trispectrum which are measured by $g_{\rm NL}$ and $\tau_{\rm NL}$ respectively. Here we only consider the case in which the curvature perturbation is originated by single source and thus $\tau_{\rm NL}=({6\over 5}f_{\rm NL})^2$. The constraint on $\tau_{\rm NL}$ from PLANCK is 
\m
\tau_{\rm NL}<2800 \quad \hbox{at $68\%$ CL} 
\n
which is much looser compared to that from $f_{\rm NL}$. The constraints on $g_{\rm NL}$ is hopefully to be done in the near future.

In \cite{Ade:2013rta} the curvaton model with quadratic potential was discussed. Here we consider a curvaton model with a polynomial potential which has been widely discussed in \cite{Sasaki:2006kq,Huang:2008zj,Enqvist:2005pg,Enqvist:2008gk,Huang:2008bg,Enqvist:2009ww,Byrnes:2011gh}. In Sec.~2, we will investigate the constraint on such a model and then figure out the prediction of $g_{\rm NL}$ in the constrained curvaton model. More discussion is given in Sec.~3.

\section{$g_{NL}$ in the constrained curvaton model}

In this section we focus on the curvaton model with a polynomial potential, as the simplest extension to the model with quadratic potential,
\m
V(\sigma)=\half m^2\sigma^2+\lambda m^4\({\sigma\over m}\)^n, 
\n
here $m$ is the mass of curvaton and $\lambda$ is the dimensionless coupling constant. Here $m^2$ is positive, but $\lambda$ can be positive or negative. For simplicity, the size of the self-interaction term compared to the mass term is characterized by 
\m
s=2\lambda \({\sigma_*\over m}\)^{n-2}. 
\n
In this paper the subscript $*$ denotes that the quantity is evaluated at the time of horizon exit of relevant perturbation mode during inflation. The equation of motion of curvaton field $\sigma$ after inflation is given by 
\m
\ddot \sigma+{3\over 2t}\dot \sigma=-m^2 \sigma\[1+n\lambda \({\sigma\over m}\)^{n-2}\].
\n
The correction from the self-interaction term in the above equation is small if $|s|\ll 2/n$.

Once the self-interaction term is taken into account, the non-Gaussianity parameters are significantly modified due to the non-linear evolution of curvaton field. In order to make our paper complete, we directly quote the results from \cite{Huang:2008bg}. 
The amplitude of primordial scalar power spectrum generated by curvaton is 
\m
P_{\zeta,\sigma} ={q^2\over 9\pi^2}r_D^2 \({H_*\over \sigma_*}\)^2, 
\n
and the non-Gaussianity parameters are given by 
\m
f_{\rm NL}={5\over 4r_D}(1+h_2)-{5\over 3}-{5r_D\over 6},
\label{fnl}
\n
and 
\m
g_{\rm NL}={25\over 54}\[{9\over 4r_D^2}(h_3+3h_2)-{9\over r_D}(1+h_2)+\half (1-9h_2)+10 r_D+3r_D^2\], 
\label{gnl}
\n
where
\m
r_D= {3\Omega_{\sigma,D}\over 4-\Omega_{\sigma,D}},
\n
$\Omega_{\sigma,D}$ is the fraction of the curvaton energy density in the energy budget at the time of curvaton decay, 
\m
q&=&{w(x_0)+n(n-1)g(n,x_0)s/2\over w(x_0)+n g(n,x_0)s/2},\\
h_2&=&{w(x_0)+n g(n,x_0)s/2\over (w(x_0)+n(n-1)g(n,x_0)s/2)^2}n(n-1)(n-2)g(n,x_0)s/2,\\
h_3&=&{(w(x_0)+n g(n,x_0)s/2)^2\over (w(x_0)+n(n-1)g(n,x_0)s/2)^3}n(n-1)(n-2)(n-3)g(n,x_0)s/2, 
\n
\m
w(x_0)=2^{1/4} \Gamma(5/4)x_0^{-1/4}J_{1/4}(x_0),
\n
\m
g(n,x_0)=&&\pi 2^{(n-5)/4}\Gamma(5/4)^{n-1} x_0^{-1/4}\nonumber \\
&&\times \[J_{1/4}(x_0)\int_0^{x_0} J_{1/4}^{n-1}(x) Y_{1/4}(x)x^{(6-n)/4}dx \right. \nonumber \\
&&\left. -Y_{1/4}(x_0)\int_0^{x_0}J_{1/4}^n(x)x^{(6-n)/4}dx\],
\n
and $x_0=mt_0=1$ denotes the time when curvaton starts to oscillate. Since $0\leq \Omega_{\sigma,D}\leq 1$, $0\leq r_D\leq 1$. The spectral index of power spectrum generated by curvaton is 
\m
n_{s,\sigma}=1-2\epsilon+2\eta_\sigma, 
\label{nss}
\n
where
\m
\eta_\sigma\equiv {V''(\sigma_*)\over 3H_*^2}={m^2\over 3H_*^2}\[1+{n(n-1)\over 2}s\]. 
\n

In the limit of $s\rightarrow 0$, the above results reduce to the model with quadratic potential and then 
\m
f_{\rm NL}={5\over 4r_D}-{5\over 3}-{5r_D\over 6},
\n
\m
g_{\rm NL}={25\over 54}\[-{9\over r_D}+\half+10 r_D+3r_D^2\]. 
\n
Considering $r_D\in [0,1]$, $f_{\rm NL}\geq -5/4$ and $g_{\rm NL}<25/12$. From the above two equations, we conclude that 
\m
g_{\rm NL}\simeq -{10\over 3}f_{\rm NL}. 
\n
In the curvaton model with quadratic potential $g_{\rm NL}$ is too small to be detected.

\begin{figure}[h]
\begin{center}
\includegraphics[width=7.2cm]{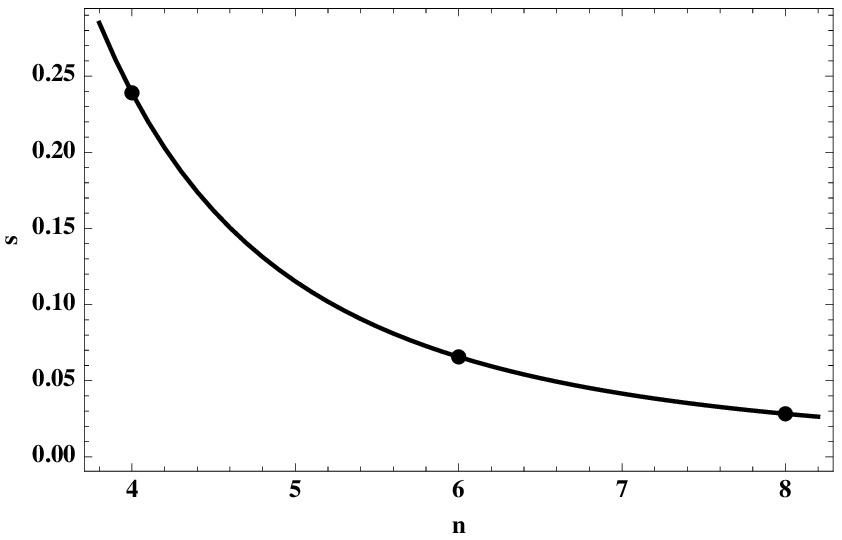}\quad \includegraphics[width=7.2cm]{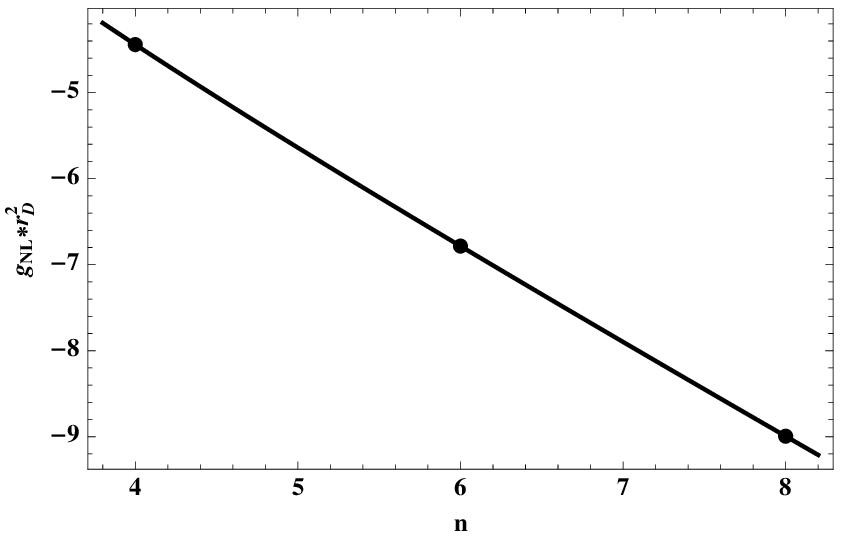}
\end{center}
\caption{The value of $s$ and $g_{\rm NL}*r_D^2$ for $f_{\rm NL}=0$ in the limit of $r_D\rightarrow 0$.  }
\label{fig:fnl0}
\end{figure}

From now on, we switch to the the case of $s\neq 0$.
Usually we may expect that $f_{\rm NL}$ should be quite large if $r_D\ll 1$. However, 
from Eq.~(\ref{fnl}), $f_{\rm NL}$ can be tuned to zero even for $r_D\ll 1$. In the limit of $r_D\rightarrow 0$, the corresponding value of $s$ for $f_{\rm NL}=0$ shows up on the left panel of Fig.~\ref{fig:fnl0}. 
Now one can check that $h_3+3h_2\neq 0$ which implies that $g_{\rm NL}$ can be arbitrarily large. The corresponding value of $g_{\rm NL}\times r_D^2$ is shown on the right panel of Fig.~\ref{fig:fnl0}. It indicates that $g_{\rm NL}$ can be negative with large absolute value. 
However the fine-tuning for $s$ is needed inevitably for obtaining a large $g_{\rm NL}$.   

\begin{figure}[h]
\begin{center}
\includegraphics[width=7cm]{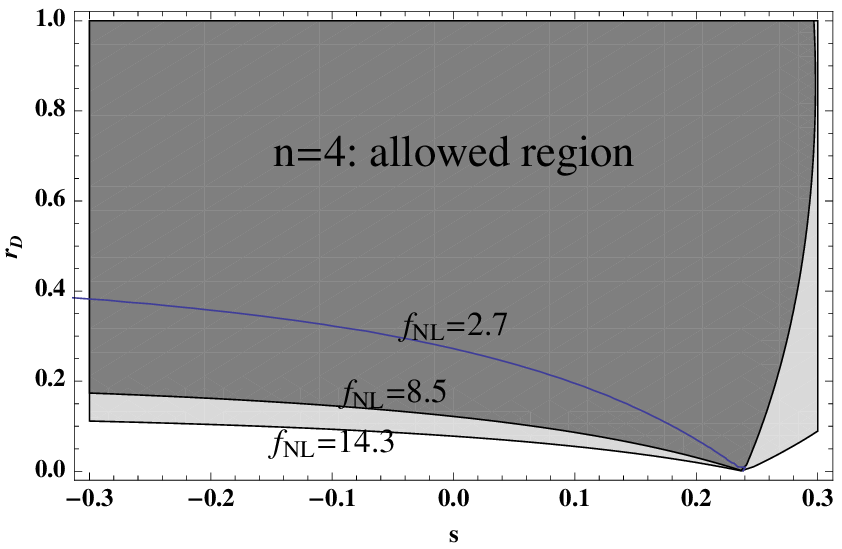} \quad
\includegraphics[width=7.3cm]{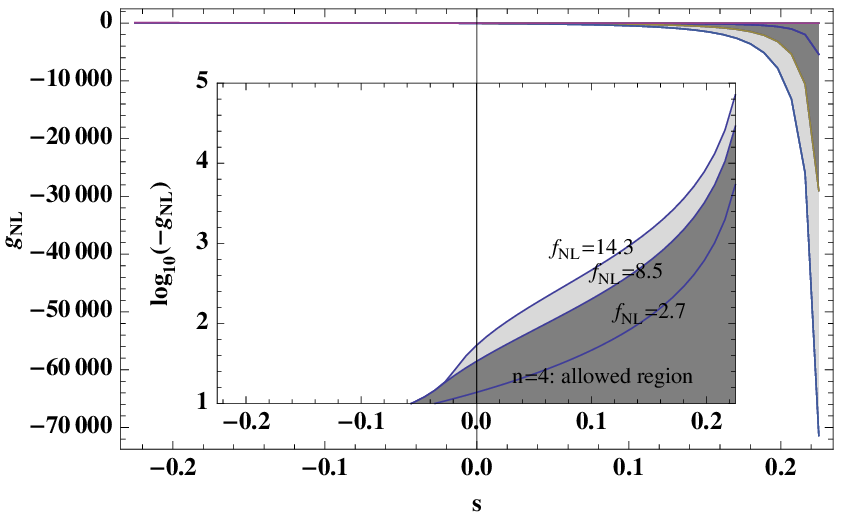}\\
\vspace{5mm}
\includegraphics[width=7cm]{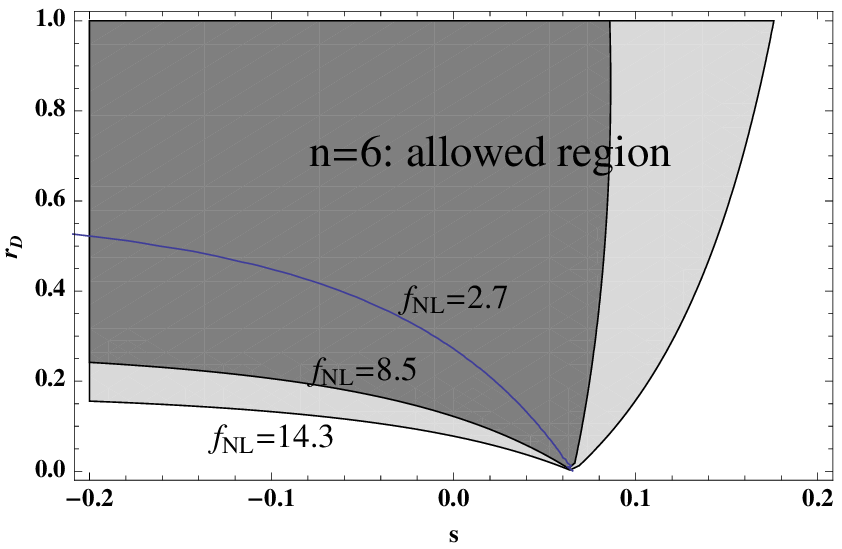} \quad
\includegraphics[width=7.3cm]{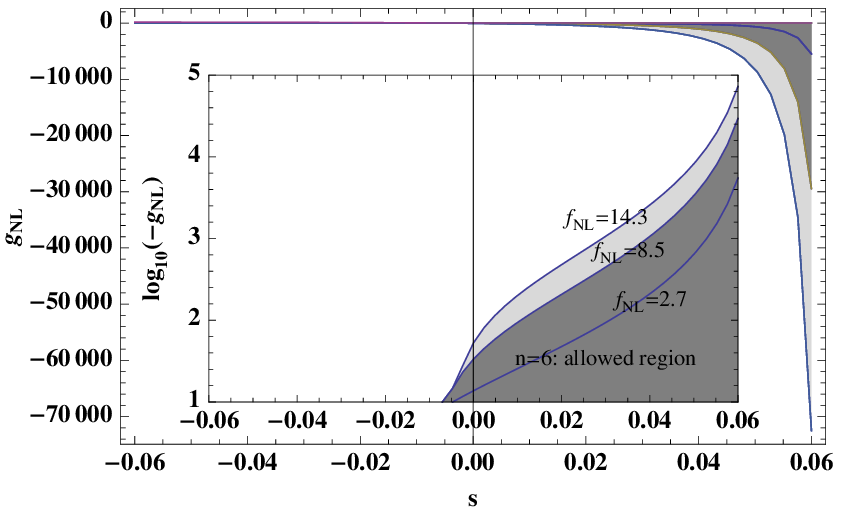}\\
\vspace{5mm}
\includegraphics[width=7cm]{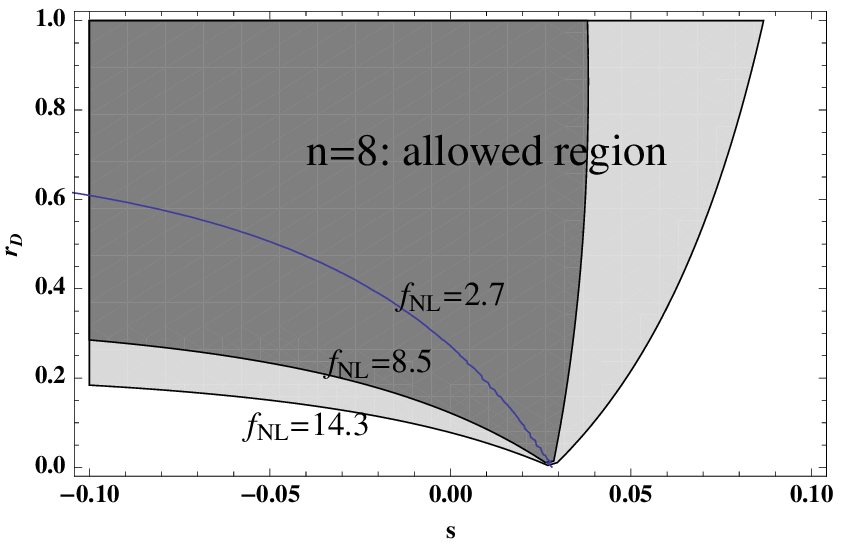} \quad
\includegraphics[width=7.3cm]{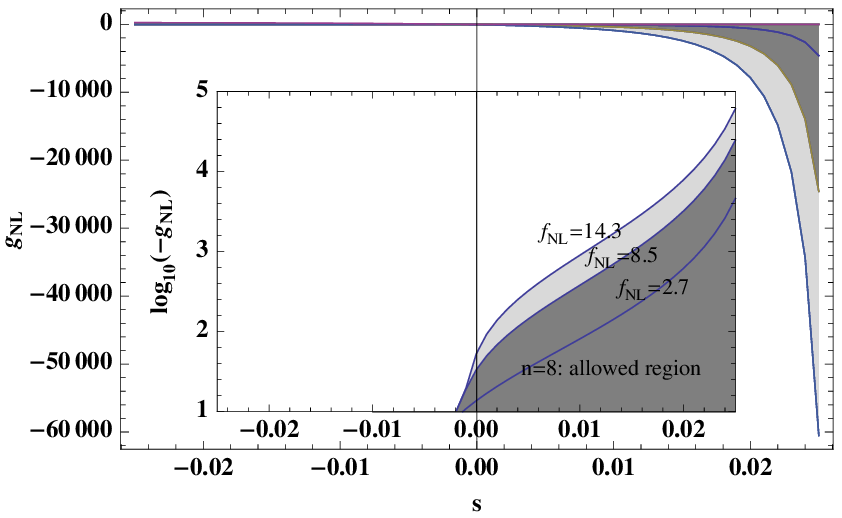}
\end{center}
\caption{The allowed parameter space and prediction of $g_{\rm NL}$ in the constrained curvaton model.  }
\label{fig:sd}
\end{figure}

Now let's go away from the fine-tuning point of $s$ in Fig.~\ref{fig:fnl0}. Since $f_{\rm NL}$ has been tightly constrained by PLANCK data in Eq.~(\ref{cfnl}), the value of $r_D$ is constrained for different value of $s$.  
See the left panel of Fig.~\ref{fig:sd}. Even though $f_{\rm NL}$ has been tightly constrained, there is still a big parameter space for the curvaton model with a polynomial potential. The allowed regions for $g_{\rm NL}$ are shown on the right panel of Fig.~\ref{fig:sd}. From the right panel of Fig.~\ref{fig:sd}, we find that the allowed value of $g_{\rm NL}$ goes to infinity when $s$ approaches to the fine-tuning point in Fig.~\ref{fig:fnl0}. 

\begin{figure}[h]
\begin{center}
\includegraphics[width=4.5cm]{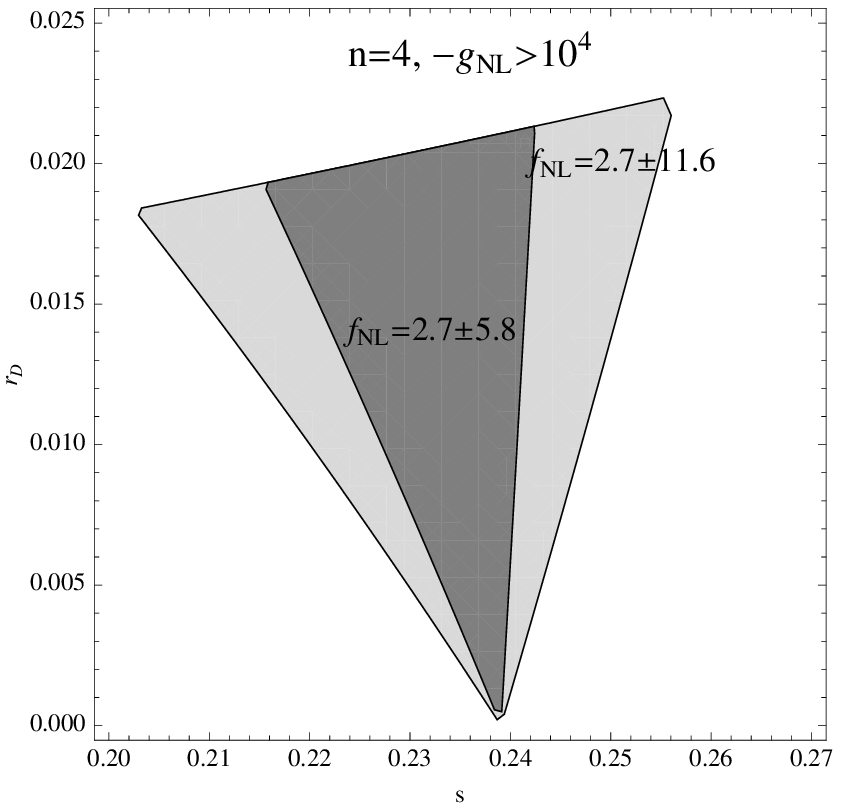}\quad
\includegraphics[width=4.5cm]{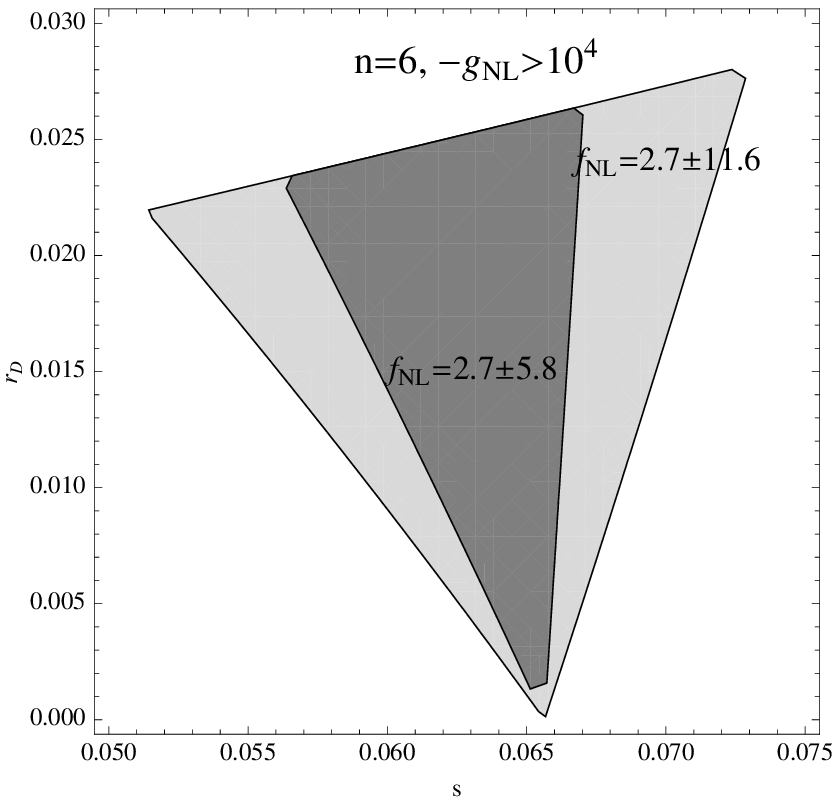}\quad
\includegraphics[width=4.5cm]{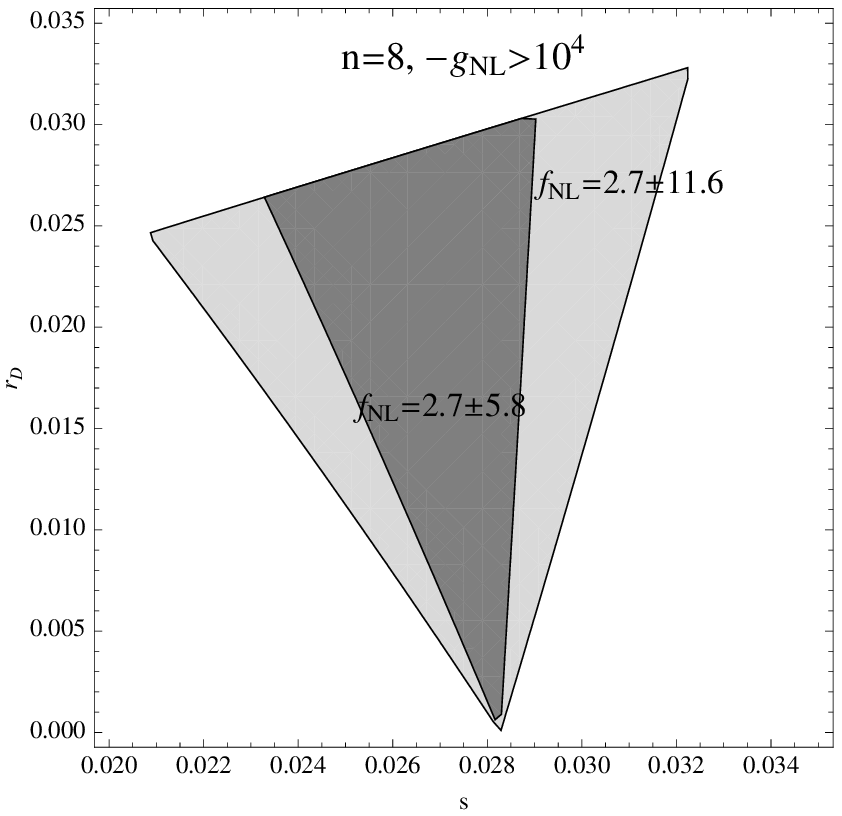}
\end{center}
\caption{The allowed parameter space for $-g_{\rm NL}>10^4$ in the constrained curvaton model.  }
\label{fig:sdg}
\end{figure}

It is also interesting for us to investigate how much fine-tuning is needed for getting a large value of $g_{\rm NL}$. For example, the parameter space for $-g_{\rm NL}>10^4$ shows up in Fig.~\ref{fig:sdg}. It implies that a large $-g_{\rm NL}$ can be obtained by around $10\%$ tuning around the fine-tuning point in Fig.~\ref{fig:fnl0}. 

Previously we mainly focus on the curvaton model with $s\ll 1$ which implies that the curvaton energy density is dominated by its mass term. One may expect that a large $g_{\rm NL}$ can be achieved if $s\gtrsim 1$. However in such a case, $f_{\rm NL}$ becomes large as well, and we have $g_{\rm NL}\sim {\cal O}(f_{\rm NL}^2)$. See \cite{Huang:2008zj,Enqvist:2009ww,Byrnes:2011gh} where the analytical and/or numerical calculations are presented in detail. Since $f_{\rm NL}$ has been tightly constrained by Planck, we conclude that $g_{\rm NL}$ cannot be quite large in the curvaton model with dominant self-interaction term.

Before closing this section, we also want to pay attention to the spectral index of power spectrum. From Fig.~\ref{fig:sdg}, the parameter $s$ needs to be positive (around the fine-tuning point) in order to achieve a large value of $g_{\rm NL}$, or numerically $\eta_\sigma/(m^2/H_*^2)\simeq 0.8,\ 0.7,\ 0.6$ for $n=4,\ 6,\ 8$ respectively. It implies $\eta_\sigma>0$ which makes the power spectrum bluer. Now one may worry about the constraint on the spectral index in Eqs.~(\ref{nspre}) and (\ref{nsp}): a red tilted power spectrum is preferred at more than $5\sigma$ level. In order to explain such a red tilted power spectrum, one need a large value of $\epsilon$ which can be realized in the inflation model with potential $U(\phi)\sim \phi^p$. 
\footnote{In \cite{Cheng:2013iya,Ade:2013rta}, the inflation model with $p>2$ is disfavored at more than $95\%$ CL because the predicted tensor-to-scalar ratio is much bigger than the bound in Eq.~(\ref{rp}). However we have to point out that the power spectrum is assumed to be completely generated by the quantum fluctuation of inflaton field $\phi$ in \cite{Cheng:2013iya,Ade:2013rta}. But here the story is totally different: the power spectrum is assumed to be generated by the curvaton field. Denoting $\beta\equiv P_{\zeta,\sigma}/P_{\zeta,obs}$, $r=16(1-\beta)\epsilon\ll 0.1$, even for $p=4$, if $(1-\beta)\lesssim {\cal O}(0.1)$ which means that the curvature perturbation is mainly generated by curvaton field. On the other hand, this example tells us that some inflation models tightly constrained by PLANCK data might be relaxed in the curvaton scenario. }
In such an inflation model, $\epsilon={p\over 4N}$ where $N$ is the number of e-folds before the end of inflation. For $N=50$, $\epsilon=0.005 p$. For $p=4$ and $m/H_*\ll 0.1$, the spectral index in the curvaton model is $n_s\simeq 0.96$ which can fit the data very well.

\section{Discussion}

In this paper we focus on the curvaton model with a polynomial potential. We find that the value of $f_{\rm NL}$ can be tuned to zero as long as the curvaton self-interaction term has suitable size compared to its mass term even when $r_D\ll 1$, and then $g_{\rm NL}$ can be arbitrarily large. However, in the curvaton model with dominant self-interaction term, this phenomenology does not happen and then $g_{\rm NL}$ cannot be quite large compared to $f_{\rm NL}^2$. A fine-tuning to the strength of curvaton self-interaction is needed for obtaining a small $f_{\rm NL}$ when $r_D\ll 1$. Our numerical analysis implies that once such a fine-tuning is abandoned, $g_{\rm NL}$ can still be large if the strength of curvaton self-interaction is not far away from the fine-tuning point. We also notice that the parameter $s$ must be positive at the fine-tuning point. It implies that the axion-type curvaton model cannot achieve a large $g_{\rm NL}$ when the constraint on $f_{\rm NL}$ from Planck is considered. To summarize, even though the curvaton model with a polynomial potential has been constrained by PLANCK data, a large positive value of $-g_{\rm NL}$ can be generated without fine-tuning.

Actually in many other models, such as the non-Gaussianity generated at the end of multi-field inflation model \cite{Huang:2009vk} and the general single-field ultra-slow-roll inflation model \cite{Huang:2013oya} and some others in \cite{Suyama:2013nva}, $g_{\rm NL}$ is an independent parameter which is not related the value of $f_{\rm NL}$ at all. 
In such kind of model, a large $g_{\rm NL}$ is still expected even though $f_{\rm NL}$ has been tightly constrained to be around zero.

Finally we want to point out that the canonical single-field slow-roll inflation predicts $f_{\rm NL}={5\over 12}(1-n_s)$ \cite{Maldacena:2002vr}. For $n_s=0.9603$, $f_{\rm NL}=0.0165$ which is still far from the current sensitivity of detector. Checking this consistency is an important task in the future.

\vspace{1.cm}

\noindent {\bf Acknowledgments}

QGH is supported by the project of Knowledge Innovation Program of Chinese Academy of Science and a grant from NSFC (grant NO. 10821504).




\end{document}